\documentclass[floatfix, a4paper, amsfonts, amssymb, amsmath, reprint, showkeys, nofootinbib, twoside, superscriptaddress, aps, pra]{revtex4-2}

\usepackage{graphicx} 
\usepackage{physics}
\usepackage{mathtools}
\usepackage{nicefrac}
\usepackage{color, colortbl} 
\usepackage[normalem]{ulem} 
\usepackage{float}
\usepackage{xr}
\usepackage{array}
\usepackage[T1]{fontenc}
\usepackage{notes2bib}
\usepackage{xurl}
\usepackage{anyfontsize} 

\begin{document}

\title{Nanosecond wavefront shaping to focus through agitated turbid media}

\author{Hugo Lassiette}
\affiliation{DOTA, ONERA, Université Paris-Saclay, 92320 Châtillon, France }

\author{Léa Krafft}
\affiliation{DOTA, ONERA, Université Paris-Saclay, 92320 Châtillon, France }

\author{Geoffrey Maulion}
\affiliation{DOTA, ONERA, Université Paris-Saclay, 92320 Châtillon, France }

\author{Jérôme Henrion}
\affiliation{DOTA, ONERA, Université Paris-Saclay, 92320 Châtillon, France }

\author{Julien Houy}
\affiliation{DOTA, ONERA, Université Paris-Saclay, 92320 Châtillon, France }

\author{Yann Lucas}
\affiliation{DOTA, ONERA, Université Paris-Saclay, 92320 Châtillon, France }

\author{Laurent Lombard}
\affiliation{DOTA, ONERA, Université Paris-Saclay, 91123 Palaiseau, France  }

\author{Bastien Rouzé}
\affiliation{DOTA, ONERA, Université Paris-Saclay, 91123 Palaiseau, France }

\author{Vincent Michau}
\affiliation{DOTA, ONERA, Université Paris-Saclay, 91123 Palaiseau, France }

\author{Willem L. Vos}, 
\affiliation{Institut Langevin, ESPCI Paris, PSL University, CNRS, 1 rue Jussieu, 75005 Paris, France}
\affiliation{Complex Photonic Systems (COPS) Chair, Faculty of Science and Technology, University of Twente, 7500 AE Enschede, The Netherlands} 
\affiliation{COPS at Photonic and Semiconductor Nanostructures (PSN) Chair, Applied Physics and Science Education (APSE) Department, Eindhoven University of Technology, Eindhoven, The Netherlands}

\author{Sébastien Popoff} 
\affiliation{Institut Langevin, ESPCI Paris, PSL University, CNRS, 1 rue Jussieu, 75005 Paris, France}

\author{Serge Meimon}
\affiliation{DOTA, ONERA, Université Paris-Saclay, 92320 Châtillon, France }
\email{Corresponding author: serge.meimon@onera.fr} 





\date{6 March 2026}


\begin{abstract}
    Multiple scattering rapidly scrambles optical fields in fog, snow and turbid water, causing optimized wavefront corrections to become obsolete on microsecond timescales. 
    Although wavefront shaping enables focusing through static scattering layers, closed-loop control in dynamically evolving media has remained experimentally challenging because the correction bandwidth must approach the intrinsic decorrelation rate of the speckle. 
    Here, we demonstrate closed-loop wavefront shaping with 32 independent degrees of freedom in an agitated turbid medium exhibiting sub-microsecond decorrelation. The medium thickness exceeds the transport mean free path, meaning the far-field speckle autocorrelation is limited to a diffraction-sized grain. Despite the microsecond decorrelation and this multiple-scattering regime, stable focusing is maintained as the correction cycle approaches the intrinsic dynamics of the medium. These results establish an experimentally accessible regime for coherent wave control in rapidly evolving complex media.
\end{abstract}
\maketitle

In strongly scattering media such as fog or turbid water, multiple scattering rapidly scrambles the optical field, causing optimal wavefront corrections to become obsolete on microsecond timescales~\cite{Pine1990JPhys, Goodman2007book}. 
While wavefront shaping enables focusing through static scattering layers~\cite{Vellekoop2007OptLett, Mosk2012NP, Rotter2017RMP, Gigan2022JPhys}, extending closed-loop control to media evolving on microsecond timescales has remained experimentally challenging~\cite{Tzang2019NP, Cheng2023NP, conkey-2012, Mididoddi2025NatPho, Feldkhun2019Opt, Liu2015NatCom}. 
Here we demonstrate closed-loop wavefront shaping with 32 independent degrees of freedom in a dynamically scattering medium exhibiting sub-microsecond decorrelation. 

Matrix-based approaches provide a complementary strategy by measuring the input–output transmission operator of the medium~\cite{Popoff2011PRL, Aubry2009AO, Rotter2017RMP, Cao2022NatPhys, Badon2020SA, Lee2023NatCom}. While such methods offer complete linear characterization in static samples, they require complete measurement of the matrix before correction, restricting their applicability in rapidly evolving media \cite{Rotter2017RMP, balondrade-2024}. In contrast, the present implementation performs direct sequential analog optimization in closed loop, requiring only the degrees of freedom necessary to generate a focus and remaining compatible with microsecond-scale decorrelation.

For characteristic scatterer velocities on the order of 0.1 m s$^{-1}$, decorrelation times in strongly scattering media fall below the microsecond  \cite{Nishimura2014JGRA, Kunkel1984JCAM}. 
Reaching this regime is not a matter of incremental acceleration, but of matching the correction bandwidth to the intrinsic evolution rate of the medium.

In the present paper, we perform an experiment at a wavelength of 1.55~$\mu$m in an agitated turbid medium with controlled flow conditions, allowing the decorrelation time to be tuned in the sub-microsecond range. The sample thickness exceeded the transport mean free path ($L > \ell_t$), placing the experiment in a multiple-scattering regime \cite{Ntziachristos2010NatMet}. The spatial autocorrelation of the transmitted far-field speckle was therefore limited to a diffraction-sized grain, indicating the absence of exploitable long-range angular memory effects \cite{Freund1988PRV, Feng1988PRL}. The optical thickness also exceeded the vanishing-distance regime \cite{Lassiette2026TBS}, ensuring that focusing could not rely on filtering residual ballistic component  \cite{Dunsby2003JPhys}.

\begin{figure}
\centering
\label{fig:figure1}
\includegraphics[width=\linewidth]{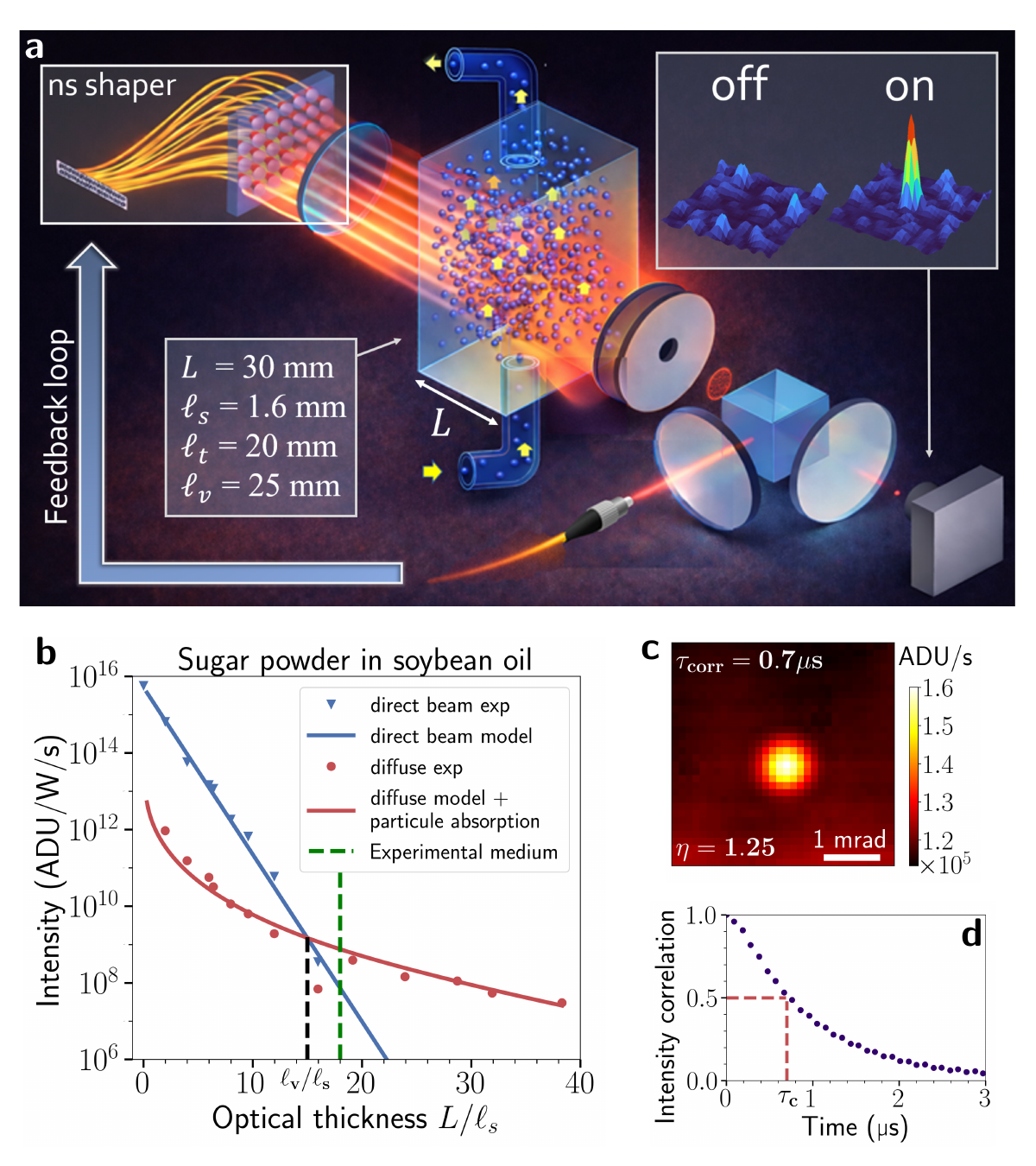}
\caption{\textbf{Operation regime and demonstration of closed-loop focusing beyond the vanishing distance.}
(a) Experimental platform implementing 32 parallel phase-modulated channels at $1.55~\mu$m. 
The channels are recombined to synthesize the incident wavefront illuminating the turbid sample; the transmitted intensity is detected in the far field.
(b) Ballistic and diffuse flux per optical mode as a function of optical thickness, defining the vanishing distance $\ell_v$. 
The experiment operates at $L \gtrsim \ell_v$, beyond the regime where ballistic contributions dominate.
(c) Intensity distribution at maximum agitation, demonstrating maintained focusing in the sub-microsecond decorrelation regime.
(d) Measured temporal autocorrelation of the transmitted speckle under dynamic conditions, defining the microsecond-scale decorrelation time.}
\end{figure}

The experimental platform implements parallel phase modulation using Electro-optic modulators with a 100 MHz bandwidth across 32 independent guided channels, whose outputs are recombined to synthesize the incident wavefront, see Fig.~1(a). 
A frequency-tagged closed-loop scheme retrieves the contribution of each degree of freedom from a single photodetected signal after an analog filter with a 1.5 MHz bandwidth, enabling simultaneous optimization within the decorrelation time~\cite{Omeara1977JOS, Shay2004SPIE}. Under quasi-static conditions, the system reaches intensity enhancements $\eta\approx 20$, consistent with the optimal enhancement value $\eta_{opt} = \frac{\pi}{4}(N_{dof} - 1) = 24.3$ reachable for iterative phase-only optimization across $N_{dof} = 32$ degrees of freedom \cite{Vellekoop2007OptLett}. 
When a flow is induced in the medium and the decorrelation time reduced to the microsecond range, stable focusing is maintained, although the enhancement progressively decreases as the correction cadence approaches the intrinsic dynamics of the medium, see Fig.~1(b,c).

\begin{figure}
\centering
\includegraphics[width=\linewidth]{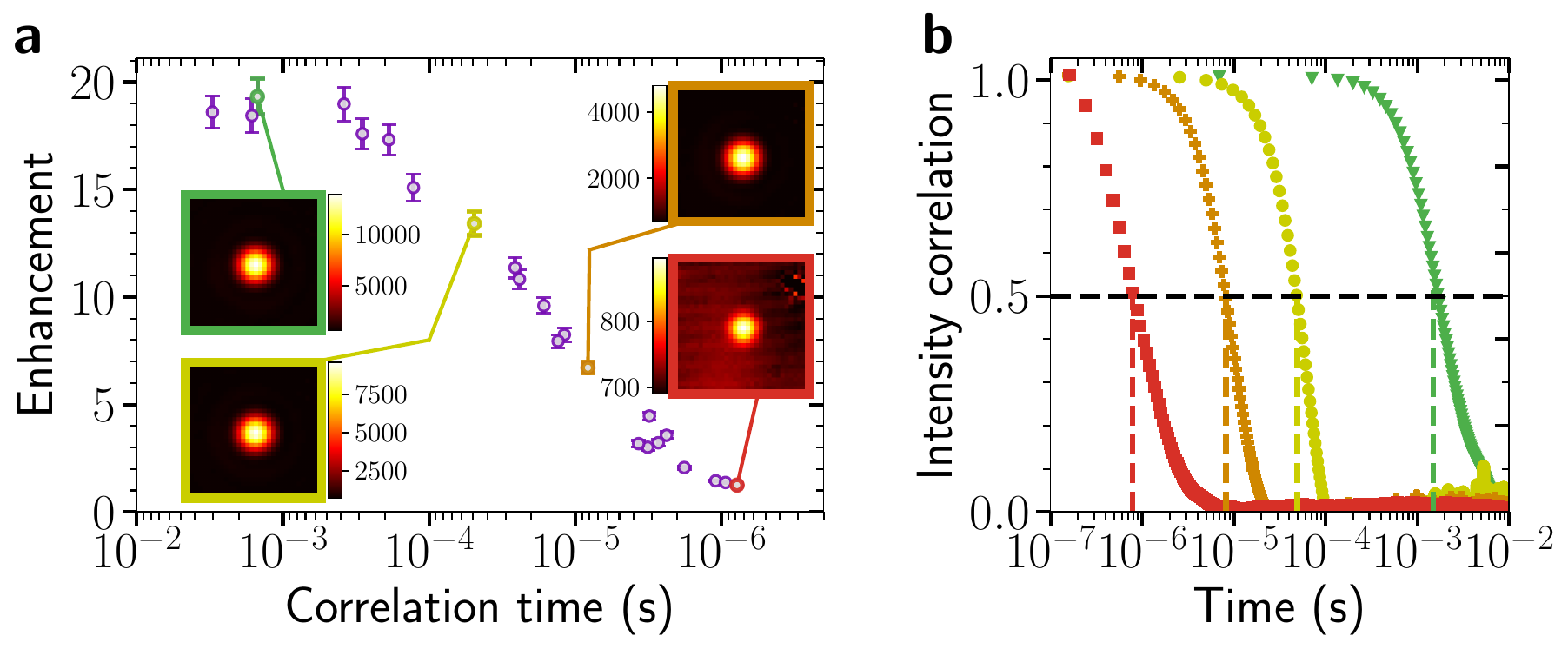}
\caption{\textbf{Focusing performance versus speckle decorrelation time.}
(a) Measured intensity enhancement as a function of decorrelation time, tuned via controlled flow velocity. 
Enhancement decreases progressively as the decorrelation time approaches the correction cycle duration.
(b) Measured temporal autocorrelation of the transmitted speckle under dynamic conditions for different time scales with caracteristic times ranging from 1 ms to 1 µs}
\end{figure}

To quantify the stability of the correction in dynamic conditions, we measured the focusing performance as a function of the speckle decorrelation time, tuned by adjusting the flow velocity (Fig.~2a). The enhancement decreases progressively as the decorrelation time approaches the correction timescale, revealing a smooth transition rather than an abrupt breakdown of the optimization process. Even for decorrelation times below one microsecond, which is close to the loop update time, a measurable and stable focus is maintained.

\begin{figure}
\centering
\includegraphics[width=\linewidth]{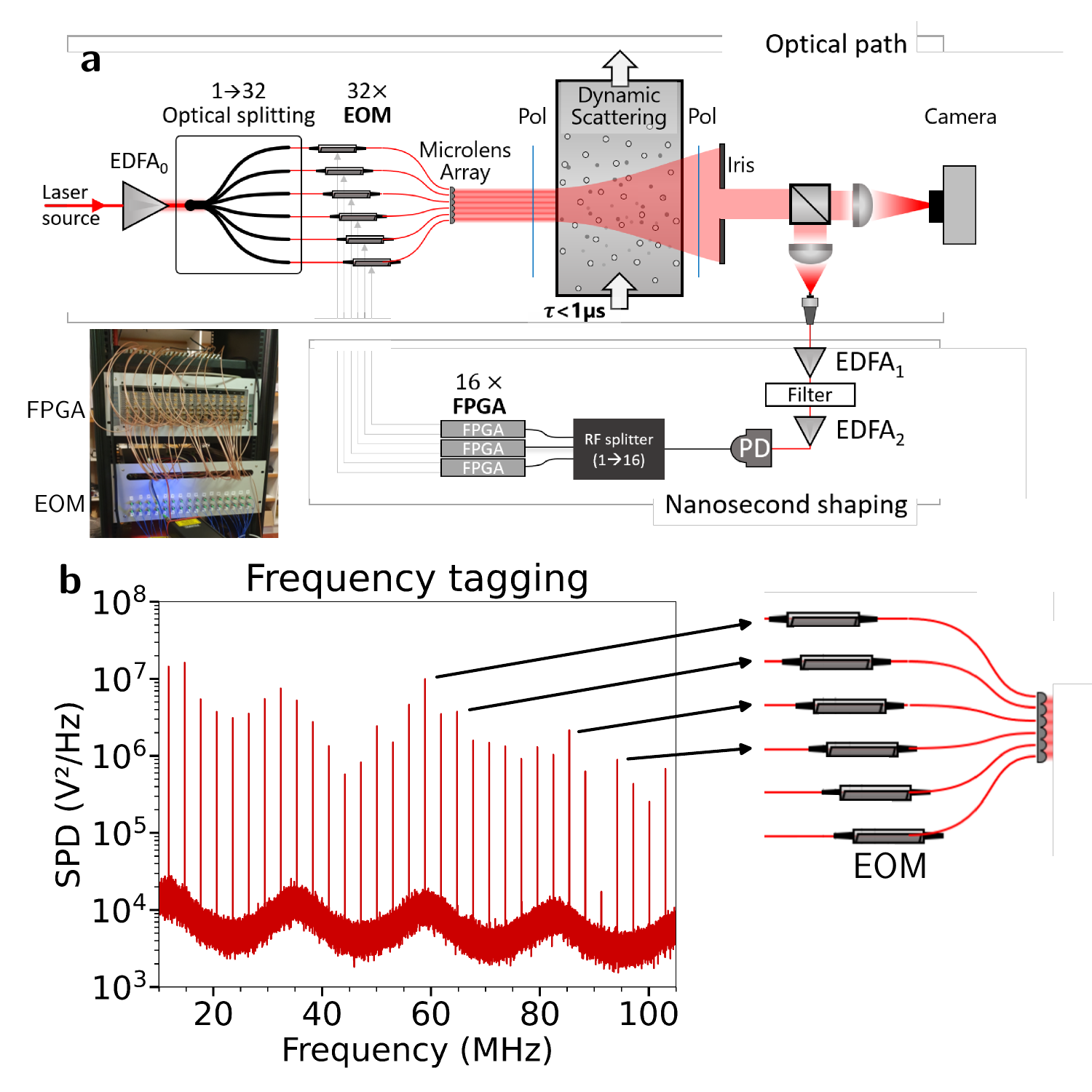}
\caption{\textbf{Parallel frequency-tagged closed-loop control.}
(a) Principle of frequency tagging applied to the 32 independently modulated phase channels.
(b) Frequency-domain spectrum of the detected intensity signal, showing resolved peaks associated with each controlled degree of freedom and enabling simultaneous gradient extraction within a sub-microsecond correction cycle.}
\end{figure}

Taken together, these observations establish that closed-loop wavefront shaping can remain effective when the correction timescale approaches the intrinsic decorrelation time of a dynamically scattering medium. Rather than relying on ballistic filtering \cite{Dunsby2003JPhys} or static matrix inversion \cite{Popoff2011PRL}, the system operates directly in a regime where multiple scattering dominates and the optical field evolves on sub-microsecond timescales. By demonstrating stable focusing under controlled agitation conditions representative of realistic flow velocities, this work defines an experimentally accessible regime for coherent wave control in rapidly evolving complex media.

\section*{Methods}

\textbf{Control architecture.}
The control architecture underlying this performance is detailed in Fig.~3. Each guided channel is phase-modulated at a distinct tagging frequency, allowing its contribution to be demultiplexed in the frequency domain \cite{Omeara1977JOS}. The resulting spectrum exhibits well-resolved peaks associated with the 32 independent degrees of freedom (Fig.3b), confirming parallel operation of the closed-loop scheme. The full correction cycle is completed on a sub-microsecond timescale, enabling the optimization process to track the medium dynamics when the decorrelation time approaches one microsecond. The architecture relies on mature guided-optics components operating at 1.55$\mu$m \cite{balasiano-2024, Brignon2013}, without requiring high-energy pulsed sources or time-gating strategies.

\textbf{Experimental platform.}
The optical source operated at 1.55~$\mu$m in continuous-wave regime. The incident wavefront was synthesized from 32 independently phase-modulated guided channels, recombined in free space to illuminate the scattering sample. Phase modulation was implemented using high-bandwidth electro-optic modulators. The transmitted intensity was collected by a single photodetector positioned in the far field.

\textbf{Scattering medium and flow control.}
The turbid medium was characterized by a scattering mean free path $\ell_s=1.6$ mm and anisotropy factor $g \approx 0.92$, leading to a transport mean free path $\ell_t = \ell_s/(1-g) = 20$ mm. The sample thickness satisfied $L > \ell_t$. Controlled agitation was achieved through regulated flow conditions, allowing systematic tuning of the speckle decorrelation time into the sub-microsecond range.

\textbf{Closed-loop optimization.}
Each channel was modulated at a distinct tagging frequency. The photodetected signal was demultiplexed in the frequency domain to retrieve the phase gradient associated with each degree of freedom. Phase updates were computed in real time and applied iteratively. The effective correction cycle duration was below one microsecond.

\textbf{Speckle correlation measurements.}
Spatial autocorrelation of the transmitted far-field speckle was computed under quasi-static conditions from temporally sampled intensity maps. The correlation width was found to be limited to a diffraction-sized grain.

\bibliographystyle{plain} 
\bibliography{_fast_wfs_biblio}

\end{document}